\documentclass[letterpaper]{aipproc}
\layoutstyle{8x11double}
\begin{document}
\title{New multi-channel electron energy analyzer with cylindrically
symmetrical electrostatic field}
\author{P. \v Ci\v zm\'ar}{address={Institute of Scientific Instruments, ASCR, Kr\'alovopolsk\'a 147, Brno, CZ-612\,64, Czech
Republic}, ,email={petr@cizmar.org}}
\author{I. Müllerov\'a}{address={Institute of Scientific Instruments, ASCR, Kr\'alovopolsk\'a 147, Brno, CZ-612\,64, Czech
Republic}, ,email={ilona@isibrno.cz}}
\author{M. Jacka}{address={formerly the University of York, Heslington, York, YO10 5DD, U.K.}, ,email={marcus@hazeshaw.com}}
\author{A. Pratt}{address={The University of York, Heslington, York, YO10 5DD, U.K.}, ,email={ap140@york.ac.uk}}

\begin{abstract}
This paper discusses an electron energy analyzer with a cylindrically
symmetrical electrostatic field, designed for rapid Auger analysis.
The device was designed and built. The best parameters of the analyzer were
estimated and then experimentally verified. 
\end{abstract}



\date{\today}

\maketitle

\section{Introduction}

One of the nearly nondestructive methods to examine surfaces of materials is the
analysis of Auger electrons. These have energies from the range roughly from 50
to 2000 eV and are emitted from the top few nanometers giving unique valuable
surface sensitivity. 
Most common sequential analyzers, such as CMA or CHA, are used. 
Although many tens of percents of emitted electrons can be collected by some
analyzers, this still may not be sufficient for fast analysis, because energies
are analyzed sequentially.  Each time the particular detection energy is
changed, there is a dead time needed by the system to get into the desired
state. Such analyzers then need much more time to obtain a spectrum. This can be
a serious issue for time dependent experiments, if the sample can be easily
damaged by the electron beam, or if a spectrum is acquired for each pixel in
an entire image. 

In general, in order to reduce the time needed to acquire a spectrum, either the
solid angle intercepted by the analyzer can be increased, or parallel detection can be employed.  It was
shown in \cite{marcus1} that it is possible to acquire the entire energy
spectrum of interest simultaneously. The basis of the analyzer used was the
two dimensional hyperbolic field \cite{marcus2}\cite{walker1}.

The approach in this work is the development of an analyzer that keeps all the
advantages of parallel acquisition and that also has a possibility to increase
the solid angle by adding cylindrical symmetry with a new focusing property.
The advantage of this solution is a further decrease of the time needed to
acquire the spectrum.

\begin{figure}
\includegraphics[width=7.5cm]{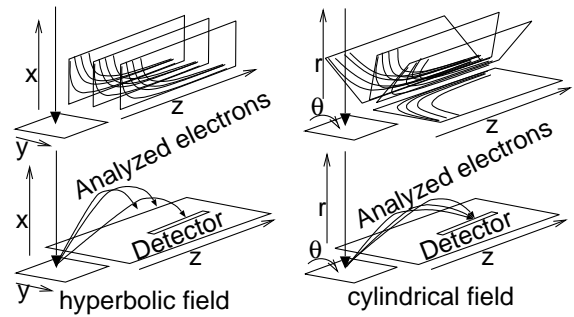}
\caption{The hyperbolic field (left-hand side) and the cylindrically symmetrical
field (right-hand side). The trajectories in the cylindrically symmetrical field 
are focused onto the detector, increasing the detected signal. Another focusing
property is added.}
\label{fig_symetrie}
\end{figure}

\section{Cylindrically symmetrical electrostatic field}
There are several conditions for the electrostatic field to be usable for
electron energy analysis. First, Laplace's equation has to be satisfied.  The
trajectories of the electrons analyzed by the field have to be focused in the
detector plane. In this case, because the field has cylindrical symmetry, there
must be the axis to axis focusing property of the field \cite{read1}. This
means that the electrons starting from one point on the axis are focused back to
the axis of symmetry, on which the detector is situated. The electrostatic field
satisfying all above conditions may be defined by the potential:

\begin{equation}
\varphi = V_0 z \log(r/R_0),
\label{art_potdef}
\end{equation}
where $V_0$ is a constant characterizing the strength of the field, $R_0$ is the
internal radius, below which there is no field, $r$ and $z$ are
cylindrical coordinates.
Knowing the potential, the equations of motion in the
cylindrical coordinate system can be written:
\begin{eqnarray}
m(\ddot r - r\dot \theta^2) &=& -qV_0z/r\nonumber\\
m(r \ddot\theta + 2\dot r\dot\theta) &=& 0\\
m\ddot z &=& -qV_0\log(r/R_0)\nonumber.
\end{eqnarray}
When the axis to axis focusing mode is employed, it can be supposed that the
particles are starting from the axis, and thus the angular component of the
velocity is zero. Then the equations of motion can be simplified to:
\begin{eqnarray}
m\ddot r &=& -qV_0zR_0/r\nonumber\\
\ddot\theta &=& 0\\
m\ddot z &=& -qV_0\log(r/R_0)\nonumber.
\end{eqnarray}
In contrast to the hyperbolic field case \cite{marcus2}, this set of differential equations
does not have an analytical solution; it has to be integrated numerically. The
trajectories were calculated using the Runge-Kutta integration method
\cite{rungekutta}.

\begin{figure}
\includegraphics[width=7.5cm]{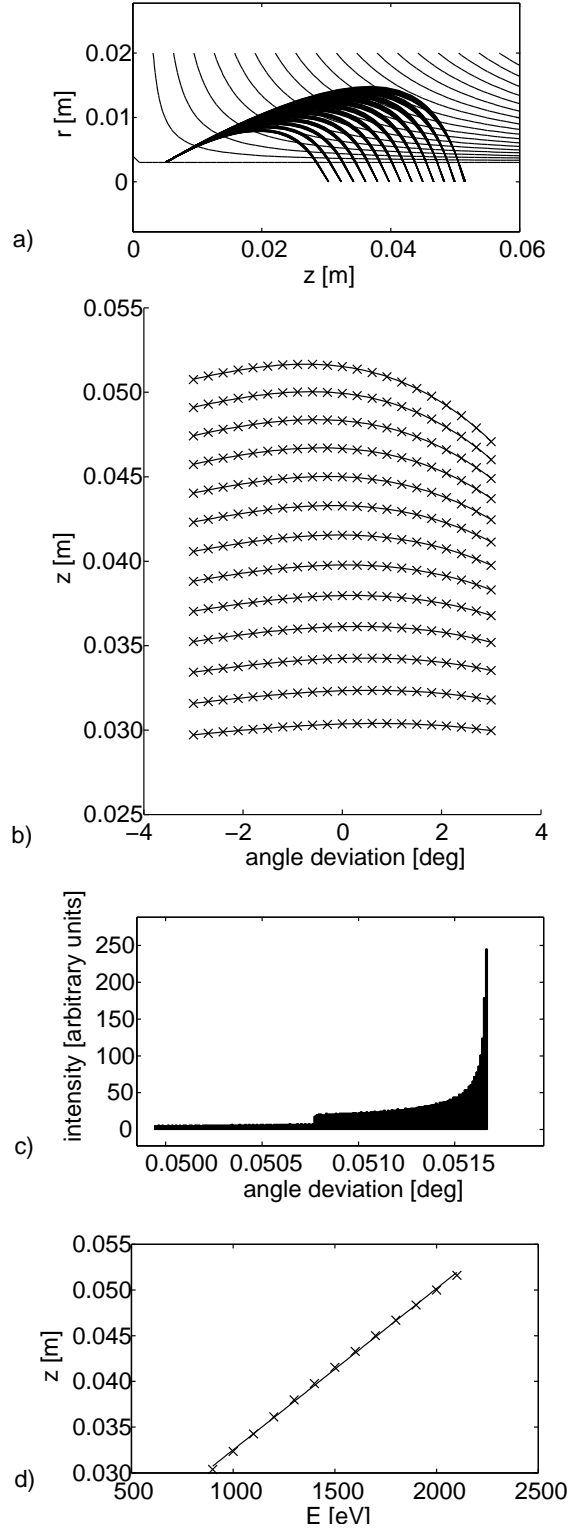}
\caption{The cylindrically symmetrical field. a) Equipotentials and trajectories
of electrons in the field for energies in the range of 900---2100 eV. The energy
step is 100 eV. Trajectories start at [0,0]m, b) Dependence of the z-coordinate
of the trajectory endpoint on the entry angle. (The presence of a maximum indicates
first order focusing), c) The point spread function (PSF) for the energy of
2100 eV d) Dependence of the z-coordinate of the endpoints on energy (integral of the
dispersion)}
\label{fig_field}
\end{figure}

It is also necessary to find the parameters of the field that produce the best
focusing and thus also the best resolution for a specified solid angle of
acceptance. One possible solution of this
problem is using a minimization algorithm. In this case for various energies
several trajectories were modeled. The sum of squares of deviations of the
endpoint positions is a satisfactory function to minimize.

From the trajectories of electrons in the field it is possible to calculate
the dispersion and the best reachable resolution of the analyzer employing this
kind of electrostatic field. 

The dispersion can be calculated just from the central trajectory for each
energy. Instead of the traditional
definition of the dispersion, $$D_r(E) = E \partial z / \partial E,$$ where the
dispersion is relative to the energy, the absolute dispersion, $$D_a(E) =
\partial z / \partial E,$$ 
is more suitable. For the CMA or other analyzers, where the detected energy is tuned, the
relative definition is more applicable, because then the dispersion is nearly
independent of energy. For the parallel analyzer the absolute definition is more
suitable for the same reason.

A similar problem is the definition of the resolution. In case of the CMA
the relative definition is used.
$$R_r(E) = \frac{E}{\Delta E(E)}$$
The reason is again the fact that this value is almost constant. 
For the same reason, in the case of the cylindrically symmetrical field analyzer or
hyperbolic field analyzer (HFA) the absolute definition is more suitable.
$$R_a(E) = \Delta E(E)$$

To be able to calculate the best possible resolution (the resolution
affected only by the properties of the electrostatic field) for each energy a
set of calculated endpoints of electron trajectories is needed.  Because the
analytical expressions of the trajectories are not known, whole trajectories have
to be calculated instead of the endpoints only. The calculation then takes more
time than in case of the HFA.

Calculation of the trajectories and their endpoints showed that for different
angles of entry different endpoints are obtained as expected. When the entry
angle is increased, the endpoint is getting farther, until a turning-point is
reached. Then the detected coordinate is decreasing. See Fig. \ref{fig_field}b.
In fact, the existence of this turning-point enables focusing (first order focusing
in this case).  The dependence of the endpoint coordinate on the angle of entry
can be very well approximated by a cubic polynomial for all energies. The
coefficients of such polynomials are then dependent on energy and can be also
very well approximated by quadratic polynomials. The detected coordinate of the
endpoint can then be expressed as
\begin{equation}
z(E,\psi) = \sum\limits_{i=0}^2\sum\limits_{j=0}^3 k_{ij}E^i\psi^j.
\label{rov_kij}
\end{equation}
The coefficients $k_{ij}$ can be calculated, and then from (\ref{rov_kij}) any
number of endpoints can be interpolated. Then it is possible to calculate
resolution from the histogram of positions, (Fig. \ref{fig_field}c) and
dependence of the endpoint position on electron energy, which is in fact
the integral of the dispersion. See Fig. \ref{fig_field}d. For the resolution, the
$\Delta z$ is defined. In this case, the density of endpoint positions (PSF
in this case) is divergent, because of the existence of turning-points. $\Delta z$ must
be defined as the distance between 20\%---80\% of the distribution function.
The $\Delta z$ varies with energy.
The calculation showed that the absolute dispersion $D_a$ is very close to a
constant. See Fig. \ref{fig_field}d. Thus 
$$
\Delta E = \Delta z/D_a.
$$

The modeling of the trajectories in the analytical field showed that the best focus
is obtained when the energies of the analyzed electrons are between 1000 eV and
2000 eV instead of the desired range of 50 eV---1000 eV, considering that the
position of the detector is given. It is possible to use 
this higher range of energies by placing an accelerator in front of the
analyzer entrance.

\section{Simulations}
To create a real analyzer with a field that has the same analytical
properties as the field defined by Eq. \ref{art_potdef}, 

electrodes of a particular shape have to be used. 
The outer shape of the analytical area of the
analyzer is determined by the geometry of the chamber, electron column and
detectors. The electrodes must
be placed where the equipotentials cross the outer shape of the analyzer. These
electrode shapes are then analytically calculated, and the shape functions are
obtained. In some cases the problem leads to transcendental equations, and
numerical methods must then be used.

To examine the behavior of the analyzed electrons in the real analyzer, which is
created by charged electrodes, simulation software can be used. 
The CPO \cite{cpo} software was suitable for this 3D pure electrostatic problem.
This software uses the boundary integral method to calculate the value of the
electrostatic potential at any particular point in the analyzer. It can also be
used to calculate the trajectories of analyzed electrons. The boundary integral
method is based on calculations of charge distributions on electrodes.
Therefore the electrode shapes had to be divided into a set of smaller
triangular or rectangular segments. 
When the charge distribution is known, electron motion within the field can be
simulated and for each electron a trajectory endpoint can be obtained. For the
endpoints obtained from CPO-3D simulation Eq. (\ref{rov_kij}) is also valid and
the coefficients may be calculated. Out of them it is possible to figure out the
resolution and the dispersion. 

\begin{figure}
\includegraphics[width=7.5cm]{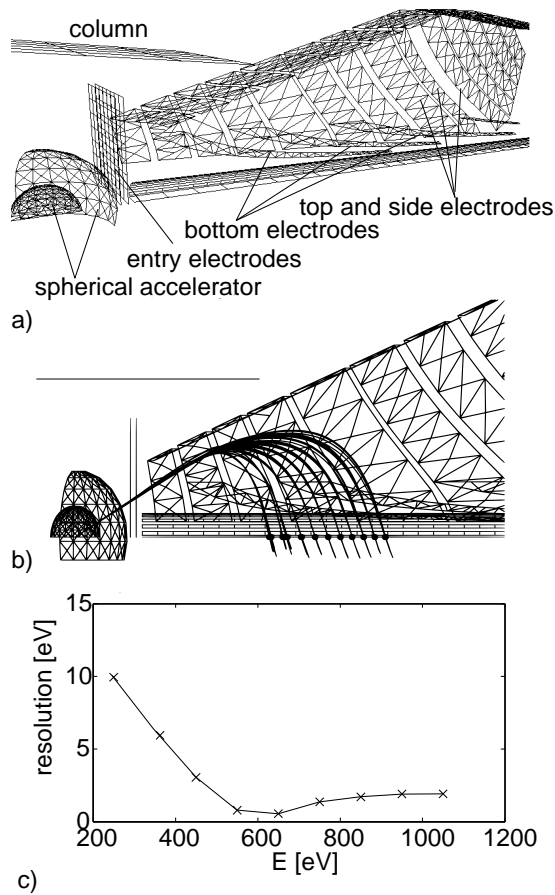}
\caption{The 3D simulation with the CPO-3D software. a) Subdivided electrodes and
description of the electrodes, b) Side view at electrodes and simulated
trajectories. The sample is positioned in the center of the hemispherical
accelerator. c) Dependence of the analyzer resolution on kinetic energy of analyzed electrons at the
sample.}
\label{fig_simulace}
\end{figure}

\section{The device}
\begin{figure}
\includegraphics[width=7.5cm]{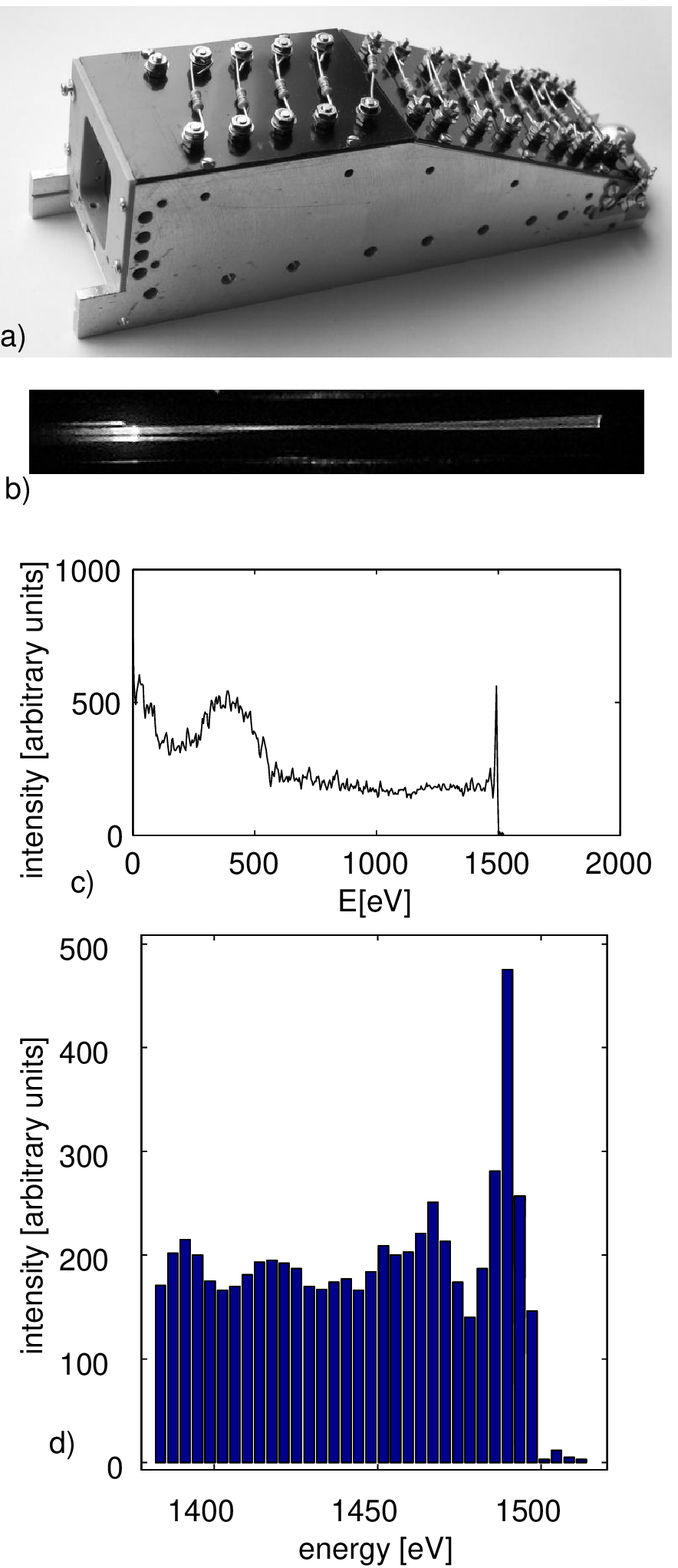}
\caption{a) Photograph of the analyzer. b) Photograph of the phosphor screen when acquiring an uncorrected electron energy
spectrum emitted from the contaminated copper sample taken at primary beam energy
of 2500 eV. c) A sum of three acquisitions of the
same spectrum. d) Detail of the relaxation peak displayed as a bar
graph.}
\label{fig_spektrum}
\end{figure}

Fig. \ref{fig_spektrum}a shows a photograph of the device.
The analyzer must satisfy several conditions to be usable in an
ultra-high-vacuum electron microscope system:
\begin{itemize}
\item The analyzer must fit into the chamber and not collide with other parts of
the microscope.\\
\item Only ultra-high-vacuum compatible materials must be used.\\
\item All parts of the analyzer have to be bakeable. They must stable at higher
temperatures.\\
\item Magnetic materials must be avoided.\\
\end{itemize}
The conditions above strongly limit the usable materials. The device
had to be designed with respect to the system used. In this case the experiment
took place in the electron microscopy laboratory at the Department of Physics of the
University of York, UK. In the past a hyperbolic field analyzer was used in this
system and the new cylindrically symmetrical field system was designed to work
in the same position in the microscope. Therefore, the new analyzer had to have
similar shape. The shapes of the electrodes were then calculated according to
this requirement. 

PEEK and Kapton materials were used for the insulating parts of the device.
Both materials are very stable at high temperatures and utra-high-vacuum
compatible. The PEEK was used to make insulating spacers. The top cover was made
of Kapton. The electrodes were accurately etched of stainless steal sheet.
The side covers and lower cylindrical electrode were made of an aluminium alloy.

A set of two concentric hemispheres was used as the accelerator, which produces
the radial electrostatic field, accelerates the electrons emitted from the sample,
and decelerates primary electrons before they land on the sample. The point
where the primary beam hits the specimen has to be in the center of the
hemispheres. As an inlet and outlet for electrons, two round holes need to be
drilled in the hemispheres. These affect the electron trajectories because they
form a lens, but the negative effect on the trajectories entering the analyzer is
significant only at the lowest energies. At higher energies this effect is negligible.

The analyzer was developed and built in the Institute of Scientific
Instruments, Academy of Sciences of the Czech Republic.

\section{Experiment}
The analyzer was used to verify the previous ideas and calculations
and to demonstrate that the cylindrically symmetrical field can be successfully
used in parallel energy analysis. The analyzer was inserted into an ultra high
vacuum system equipped with an electrostatic column. The pressure in the sample
chamber was on the order of $10^{-8}$ Pa. As a sample a piece of copper foil
was used, although the sample was not cleaned. For a detector the electrons were
multiplied by micro channel plate and then converted to light quanta with a
phosphor screen. The image on the screen was then photographed through a
vacuum window using Konica-Minolta Z3 digital camera.

\section{Conclusion}

 From the simulation, the absolute dispersion is nearly constant at $1.97\times
10^{-2} {\rm mm\cdot eV^{-1}}$ 
The resolution varies from 10 eV to 2 eV, for most energies keeps below 2 eV, which
well corresponds to \cite{read2}. The result of the experiment is a 400 pixel
long spectrum. (See Fig. \ref{fig_spektrum}.) The relaxation peak is 1 pixel wide, which shows that the energy resolution is
better than 3 eV (Fig. \ref{fig_spektrum}d) at 1500 eV, which
satisfies the theoretical estimations. The experiment also showed the
cylindrical focusing. In the photograph of the screen (Fig.
\ref{fig_spektrum}b) different widths of illuminated area can be seen. These
may be caused by a slight misalignment that occurred during bake-out.
This fact also affects the signal levels of different channels. The peak between
300 eV and 500 eV is partially caused by carbon (the main matter covering
the surface of the sample) because the sample was not cleaned.

\section{Acknowledgment}
The work was supported by the ASCR grant agency project number IAA1065304. We
also gratefully acknowledge help of Prof. B. Lencova and Mr. Pavel Klein.

\end{document}